\documentclass[letterpaper,comsoc]{IEEEtran}

\usepackage{graphicx,epsfig}
\usepackage[noadjust]{cite}
\usepackage{mcite}
\usepackage{amsfonts,helvet}
\usepackage{fancyhdr}
\usepackage{threeparttable}
\usepackage{epsf,epsfig}
\usepackage{amsthm}
\usepackage{amsmath}
\usepackage{siunitx}
\usepackage{amssymb}
\usepackage{comment}
\usepackage{dsfont}
\usepackage{subfigure}
\usepackage{color}
\usepackage[linesnumbered,noend,ruled]{algorithm2e}
\usepackage{algpseudocode}
\usepackage{algcompatible}
\usepackage{enumerate}
\usepackage{cancel}
\usepackage{bbm}
\usepackage{graphicx,subfigure}
\usepackage{graphicx}
\usepackage{amsmath}
\usepackage{booktabs}
\usepackage{array}
\usepackage{eucal}
\usepackage{microtype}

\newtheorem{remark}{Remark}

\def\bI{{\bf I}}

\def\bbR{\mbox{$\mathbb{R}$}}

\def\BibTeX{{\rm B\kern-.05em{\sc i\kern-.025em b}\kern-.08em
    T\kern-.1667em\lower.7ex\hbox{E}\kern-.125emX}}

\begin{document}

\title{Predictive Beamforming via Echo-Beam Modality Separation with Cross-Attention for ISAC Systems}

\author{Seokjun Park, {Jihwan Jang}, and Jinseok Choi
\thanks{
S. Park and J. Choi are with the Department of Electrical Engineering, Korea Advanced Institute of Science and Technology, Daejeon, South Korea (e-mail: {\texttt{\{sj.park, jinseok\}@kaist.ac.kr}}).
J. Jang is with LG Electronics, Seoul, South Korea (e-mail: {\texttt{jangjh1207@gmail.com}}).
}
}

\maketitle

\begin{abstract}
In this paper, we propose a deep learning-based predictive beamforming algorithm for integrated sensing and communications in vehicle-to-infrastructure networks, where echo signals are leveraged to enhance communication performance.
We formulate a sum spectral efficiency maximization problem under a sensing signal-to-interference-plus-noise ratio (SINR) constraint and construct a penalty-based unconstrained surrogate objective for network training.
Based on this formulation, the beamforming matrix is predicted using sensing echoes and historical beamformers from previous time slots.
To this end, we develop a predictive beamforming network that processes sensing echoes and beamforming inputs through separate encoders to preserve their modality-specific characteristics. This separate design enables each modality to learn rich and complementary representations which are more effectively exploited through a cross-attention mechanism compared to naive input fusion.
The resulting features are fused via cross-attention, while gated recurrent units capture temporal dependencies across time slots.
Simulation results validate the performance of the proposed method.
\end{abstract}


\section{Introduction}
Connected and autonomous vehicles, as a key application of beyond 5G (B5G) and 6G networks, demand ultra-reliable gigabit-level links and centimeter-level sensing accuracy to enable advanced services such as autonomous driving and intelligent transportation systems \cite{liu2022survey,charan2021vision}.
However, conventional architectures that deploy separate base stations (BSs) for communications and sensing suffer from inefficient spectrum usage and high hardware cost.
Integrated sensing and communications (ISAC) addresses these limitations by unifying both functions \cite{meng2023sensing,li2024framework,choi2024joint,park2025covariance}.

In the ISAC system, sensing echo signals can be exploited to facilitate predictive beamforming without necessitating pilot symbols \cite{liu2020radar,ying2021particle}, thereby reducing overhead for communications;
the Doppler frequency and time delay information extracted from sensing echoes reflect the kinematic state of the vehicle, enabling  tracking and prediction.
In \cite{liu2020radar}, an extended Kalman filter (EKF)-based predictive beamforming scheme was proposed to utilize sensing echoes for communications, and in \cite{ying2021particle}, a scheme based on the particle filter (PF) was developed to address the non-linearities and non-Gaussian noise in the trajectories of vehicles.
These filter-based methods typically depend on intensive matched filtering and recursive state tracking, which  incur significant processing overhead.

To reduce processing overhead and enhance performance, deep learning (DL)-based predictive beamforming methods were proposed \cite{liu2022learning,zhang2023deep,zhang2024transformer}.
For example, \cite{liu2022learning} proposed a long short-term memory (LSTM)-based algorithm that takes historical channel matrices as inputs, bypassing the tracking process.
However, relying on channel matrices still necessitates parameter estimation and channel reconstruction from the sensing echoes, and the sequential processing of the LSTM architecture restricts parallel computation, which increases both training time and real-time inference latency.

To overcome these limitations, \cite{zhang2024transformer} proposed a transformer-based method that directly generates the predictive beamforming matrix without explicit channel reconstruction, where the sensing echo and the beamformer from the previous time step are simply concatenated into a single input.
This naive fusion makes it difficult to preserve and effectively utilize the distinct characteristics of each echo and beamformer feature, and relying exclusively on a single previous step provides insufficient temporal information to capture dynamic channel variations, so that the prediction performance degrades significantly as the vehicle speed increases.
In this regard, we are motivated to develop an architecture that decouples sensing echoes and beamformers to explicitly extract their  features, while capturing the  temporal dependencies across multiple time slots  for high-mobility scenarios.

In this paper, we propose a DL-based predictive beamforming framework for vehicular ISAC networks, where sensing echoes are  exploited to elevate communication performance.
We first formulate a sum spectral efficiency (SE) maximization problem under a sensing  SINR constraint, which is then relaxed into a penalty-based unconstrained surrogate objective for DL training.
Driven by this formulation, we develop a predictive beamforming network that leverages sensing echoes and historical beamformers across previous time slots, explicitly decoupling these inputs to preserve their modality-specific characteristics.
This decoupled design extracts  mobility-related features from the echoes and prior beamforming decisions from historical beamformers.
Subsequently, a cross-attention mechanism effectively fuses these distinct features, while gated recurrent units (GRUs) capture the temporal dependencies across multiple time slots.
Simulation results demonstrate that the proposed framework consistently outperforms state-of-the-art benchmarks.

\textit{Notation}:
$\text{a}$ is a scalar, ${\bf{a}}$ is a vector, and ${\bf{A}}$ is a matrix.
The superscripts $(\cdot)^{\sf T}$, $(\cdot)^{\sf H}$, $*$, and $(\cdot)^{-1}$ denote matrix transpose, Hermitian, conjugate, and inversion, respectively.
$\mathbb{E}[\cdot]$, ${\rm Tr}(\cdot)$, and $\otimes$ represent  expectation operator, trace operator, and Kronecker product, respectively.
${\bf{I}}_K$ is the  $K \times K$ identity matrix.
${\bf 0}_{N}$ is the $N\times 1$ zero vector.
$\mathrm{Re}\{\cdot\}$ and $\mathrm{Im}\{\cdot\}$ denote the real and  imaginary parts of a complex value.

\section{System Model}
\subsection{Signal Model}
We consider vehicular networks in which the BS, equipped with a MIMO uniform linear array (ULA)  consisting of ${N}_{t}$ transmit antennas and $N_r$ receive antennas, serves $K$ user vehicles with high mobility.
We assume that the sensing target is the same as the communication user, and both the communication link and the sensing  path are  line-of-sight (LOS).
Let $n$  denote the index of time slots.
Each time slot has duration $\Delta T$, and at time  $t \in [0,\Delta T)$ within time slot $n$, the ISAC signal transmitted from the BS to the $K$ vehicles is expressed as
${\bf s}_n(t)=[s_{1,n}(t),...,s_{K,n}(t)]^{\sf T}$.
We assume the geometric parameters, i.e., the angle, Doppler frequency, time delay, and distance, remain  quasi-static and change across time slots according to the mobility model.

At the BS, ${\bf s}_n(t)$ is precoded by ${\bf F}_n$ before transmission as
\begin{align}
    {\bf{x}}_{n}(t) = \sqrt{P}{\bf F}_n{\bf s}_n(t),
\end{align}
where   ${\bf F}_n = [{\bf f}_{1,n},\cdots,{\bf f}_{K,n}] \in {\mathbb{C}^{N_t\times K}}$ denotes the downlink beamforming matrix,  ${\bf f}_{k,n} \in {\mathbb{C}}^{N_t}$ is the beamformer for vehicle $k$, and ${P}$ is the maximum transmit power.
Let the transmit and receive steering vectors be  ${\bf a}(\theta_{k,n}) = \sqrt{{1}/{N_t}}[1,e^{-j\pi\cos{\theta_{k,n}}}, \cdots, e^{-j\pi(N_t-1)\cos{\theta_{k,n}}}]^{\sf T}$ and ${\bf b}(\theta_{k,n}) = \sqrt{{1}/{N_r}}[1,e^{-j\pi\cos{\theta_{k,n}}}, \cdots, e^{-j\pi(N_r-1)\cos{\theta_{k,n}}}]^{\sf T}$, respectively, where $\theta_{k,n}$ is  the  angle between vehicle $k$ and the BS.
Then, the sensing echo signal vector at the BS reflected from $K$ vehicles in the $n$th time slot \cite{liu2020radar, liu2022learning} is represented as follows:
\begin{align}
    \!{\bf{r}}_n(t)\!=\!\sum_{k=1}^{K} \beta_{k,n}e^{j2\pi\mu_{k,n}t}{\bf b}(\theta_{k,n}){\bf a}^{\sf H}(\!\theta_{k,n}\!){\bf x}_n(t\!-\!\zeta_{k,n})\!+\!{\bf n}(t),
    \label{eq:echo}
\end{align}
where $\beta_{k,n} = \sqrt{N_t N_r}\frac{\phi}{2 d_{k,n}}$ is the reflection coefficient, $d_{k,n}$ is the distance between the BS and  vehicle $k$, $\phi$ is the complex radar cross-section,  $\mu_{k,n}$ is the Doppler frequency, $\zeta_{k,n}$ is the time delay, and ${\bf n}(t) \sim {\mathcal{CN}(0,\sigma^2\bI_{N_r})}$ is the complex additive white Gaussian noise with zero mean and variance  $\sigma^2$.
{We note that clutter from surrounding scatterers is neglected, since dominant
road-side clutter is typically stationary and can be suppressed in the Doppler domain before feature extraction \cite{richards2014fundamentals}.
Extending the proposed framework to account for the residual clutter caused by imperfect suppression and moving non-user scatterers is left as future work.}

By leveraging the massive MIMO system at the BS, the steering vectors corresponding to distinct angles become increasingly separable \cite{marzetta2016fundamentals}, which allows the BS to distinguish individual vehicles based on their angles of arrival and process them independently.
{Accordingly, the echo signal from the $k$th vehicle can be extracted from the aggregate signal in \eqref{eq:echo} by applying a spatial filtering process \cite{van2002optimum,liu2022learning} as
\begin{align}
    \label{eq:rxecho0}
    &e_{k,n}(t) = {\bf b}^{\sf H}({\hat \theta}_{k,n}){\bf  r}_{n}(t)
    \\
    \nonumber
    &\approx\beta_{k,n}e^{j2\pi\mu_{k,n}t}{\bf a}^{\sf H}(\theta_{k,n}){\bf x}_n(t-\zeta_{k,n})\!+\!{ n}_{k,n}(t)
    \\
    \label{eq:rxecho}
    &+\!\sum_{i\neq k}\!\beta_{i,n}e^{j2\pi\mu_{i,n}t}\big[{\bf b}^{\sf H}\!(\theta_{k,n}){\bf b}(\theta_{i,n})\big]{\bf a}^{\sf H}\!(\theta_{i,n}){\bf x}_n(t\!-\!\zeta_{i,n}),
\end{align}}
where ${\hat \theta}_{k,n}$ denotes an estimate of ${ \theta}_{k,n}$ obtained via angle of arrival estimation \cite{van2002optimum}, and $n_{k,n}(t) = {\bf b}^{\sf H}({\hat \theta}_{k,n}){\bf n}(t)$.
For simplification, we assume that  ${\hat \theta}_{k,n} \approx {\theta}_{k,n}$, and hence,
${\bf b}^{\sf H}({\hat \theta}_{k,n}){\bf b}({\theta}_{k,n}) \approx 1$.
We consider $L$ samples in each time slot.
Then, by aggregating $L$ samples within the time slot, we can obtain
\begin{align}
    \label{eq:ek}
    {\bf e}_{k,n} = [e_{k,n}(1),\cdots,e_{k,n}(L)]^{\sf T} \in {\mathbb{C}}^{L}.
\end{align}

Next, the received signal of vehicle $k$ at time instant $t$ within time slot $n$ can be expressed  as \cite{liu2022learning,zhang2024transformer}
\begin{align}
    &y_{k,n}(t)\!=\!\! \sqrt{P\alpha_{k,n}}e^{j2\pi\mu_{k,n}t}{\bf a}^{\sf H}(\theta_{k,n})\!\!\sum_{i=1}^{K}{\bf f}_{i,n}s_{i,n}(t) \!+\! z_{k,n}(t),
    \label{eq:Received signal}
\end{align}
where $\alpha_{k,n}$ is given by $N_t\alpha_0(d_{k,n}/d_0)^{-\eta}$.
Here, $\alpha_0$ is the path loss at a reference distance $d_0$ and $\eta$ denotes
the path loss exponent.
In addition, $z_{k,n}(t) \sim  {\mathcal{CN}(0,\sigma^2)}$ is the noise of $k$th vehicle with variance $\sigma^2$.

\subsection{Vehicle Mobility}
To support beam prediction under mobility, vehicle motion is described using a kinematic equation \cite{liu2022learning,zhang2024transformer}.
Vehicles are assumed to move parallel to the road with a nearly constant velocity direction within each time slot, and the velocity model is characterized by the velocity magnitude as follows:
\begin{align}
    \label{eq:velocity}
    v_{k,n} = v_{k,n-1}+ \Delta v_{k,n-1},
\end{align}
where $v_{k,n}$ denotes the average velocity magnitude of  vehicle $k$ and $\Delta v_{k,n-1}$ is the velocity increment
within time slot $(n-1)$.
Then, the kinematic equations of angle and distance are expressed as \cite{liu2022learning,zhang2024transformer}
\begin{align*}
    &\sin({\theta_{k,n}-\theta_{k,n-1}}) {d_{k,n}}= \sin{\theta_{k,n-1}}{v_{k,n}\Delta T},
    \\
    &d_{k,n}^2\!=\!d_{k,n-1}^2\!+\!(v_{k,n}\Delta T)^2\!-\!2d_{k,n-1}v_{k,n}\Delta T \cos{\theta_{k,n-1}}.
\end{align*}
The kinematic equations show that the mobility leads to changes in the angle and distance between the BS and vehicle over time slots, which motivates accurate beam prediction.
{For example, a variation in the vehicle speed changes both the distance and the angle between consecutive time slots. We note that the mobility model is used only to generate the training and test trajectories, and the kinematic equations are not embedded in the proposed network, unlike model-based trackers such as EKF \cite{liu2020radar}. Hence, the network learns the temporal evolution from the historical sensing echoes and beamformers, and can be retrained with trajectories from other mobility models, e.g., microscopic traffic simulators, without modifying its architecture.}

\subsection{Problem Formulation}
We formulate an optimization problem to maximize the sum SE while satisfying both the sensing SINR requirement and the total transmit power budget.
Then, the received sensing SINR \cite{liu2022learning,zhang2024transformer} of vehicle $k$  at the BS is defined as
\begin{align}
    &\xi_{k,n}({\bf F}_n) = \frac{|\beta_{k,n}|^2|{\bf a}^{\sf H}(\theta_{k,n}){\bf f}_{k,n}|^2}{|\beta_{k,n}|^2\sum_{j\neq k}^{K}|{\bf a}^{\sf H}(\theta_{k,n}){\bf f}_{j,n}|^2+\Psi +\sigma^2/P},
    \label{eq:radarSINR}
\end{align}
where $\Psi = \sum_{i\neq k}^{K}|\beta_{i,n}|^2|{\bf b}^{\sf H}(\theta_{k,n}){\bf b}(\theta_{i,n})|^2||{\bf a}^{\sf H}(\theta_{i,n}){\bf F}_n||^2$.
Accordingly, using \eqref{eq:Received signal} and \eqref{eq:radarSINR}, we formulate an optimization problem as follows:
\begin{align}
    \mathop{{\text{maximize}}}_{{\bf F}_{n}}& \;\;
   {\mathbb{E}}_{{\bf H}_n\!|{\bf  E}_n^{\tau},{\bf F}_n^{\tau}}\!\left[\!\sum_{k = 1}^{K}\log_2\!\left(\!\!1\!+\!\frac{\!|{\bf h}_{k,n}^{\sf H}{\bf f}_{k,n}|^2}{\sum_{i\neq k}^{K}\!|{\bf h}_{k,n}^{\sf H}{\bf f}_{i,n}|^2\!+\!\frac{\sigma^2}{P}}\!\right)\!\right]
    \label{eq:reformulated opt1}
    \\
    \mathop{{\text{subject to}}} \;\; &{{\rm Tr}}\left( {\bf F}_n{{\bf F}^{\sf H}_n}\right) \leq  1,
        \label{eq:TXpow}
    \\
    &\xi_{k,n}({\bf F}_n) \ge {\Gamma}_{k,n}\label{eq:SINR},
\end{align}
where ${\bf h}_{k,n}^{\sf H} = \sqrt{ \alpha_{k,n}}{\bf a}^{\sf H}(\theta_{k,n})$ is the channel vector between the BS and the $k$th vehicle and ${\bf H}_n = [{\bf h}_{1,n},\cdots,{\bf h}_{K,n}]$ is the corresponding channel matrix.
Here, the historical echo signals from time slot $(n-\tau)$ to time slot $(n-1)$ are denoted by ${\bf  E}_n^{\tau} = [ {\bf E}_{n-\tau},  {\bf E}_{n-\tau+1},\cdots, {\bf E}_{n-1}]$, where $ {\bf E}_{l}=[ {\bf e}_{1,l},\cdots, {\bf e}_{K,l}]^{\sf H}$ represents the echo signals from all $K$ vehicles at time slot $l$.
The historical beamformers are denoted by ${\bf F}_n^{\tau} = [{\bf F}_{n-\tau}, {\bf F}_{n-\tau+1},\cdots,{\bf F}_{n-1}]$, where ${\bf F}_{l}=[{\bf f}_{1,l},\ldots,{\bf f}_{K,l}]$ represents the beamforming matrix for  $K$ vehicles at time slot $l$.
Moreover, ${\Gamma}_{k,n}$ denotes the minimum sensing SINR threshold, \eqref{eq:TXpow} represents the transmit power constraint of the BS, and \eqref{eq:SINR} indicates the sensing SINR constraint.

\begin{figure}[!t]
   \centering
    \includegraphics[width=0.51\textwidth]{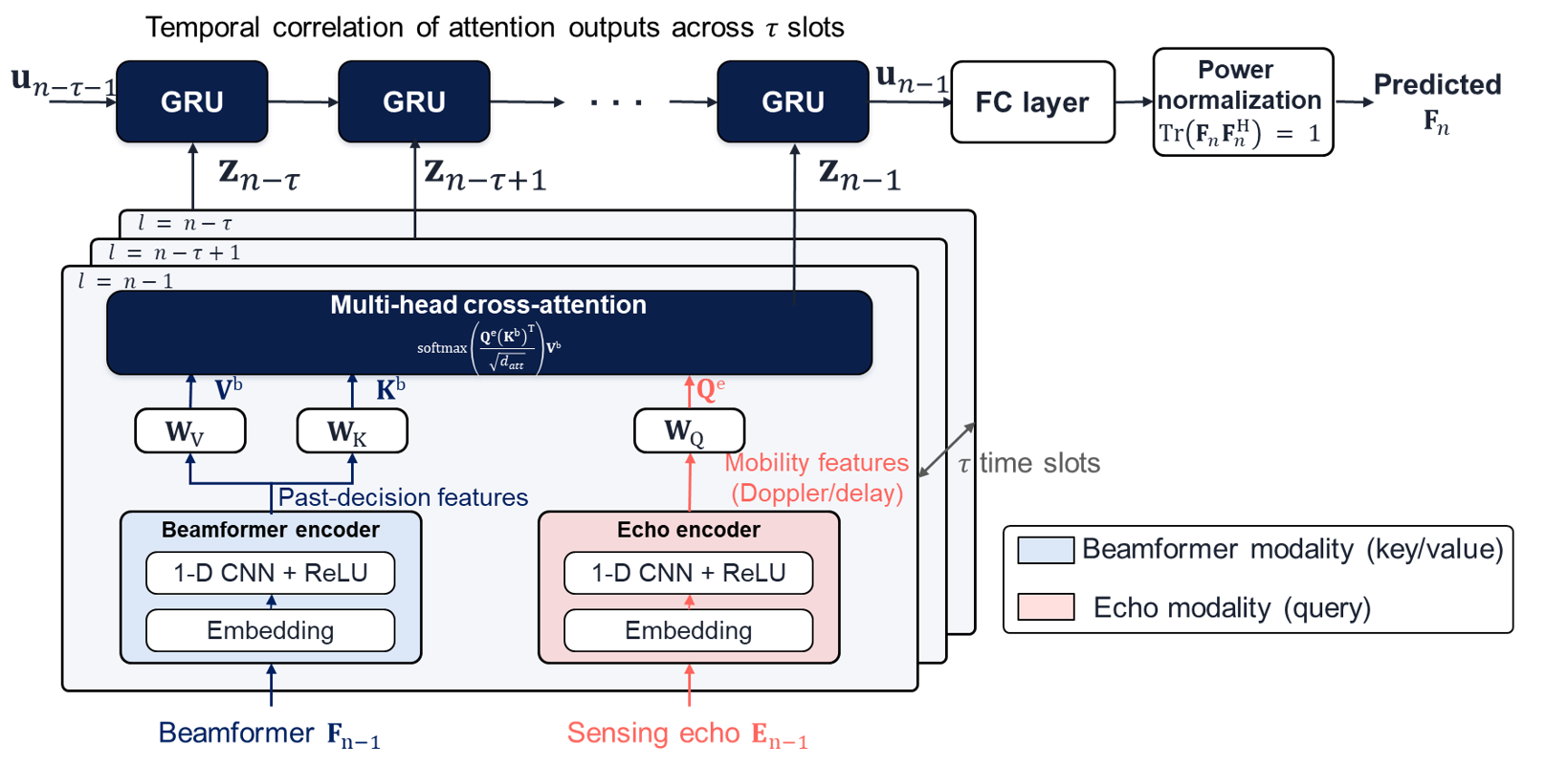} 
    \caption{Proposed predictive beamforming C-GRU-PB-Net architecture.}
    \label{fig:systemmodel}
\end{figure}
\section{Cross-attention-based Predictive Beamforming  for ISAC}
To facilitate gradient-based training within a DL framework, we enforce the transmit-power constraint by normalizing the network output and incorporate the sensing SINR constraint through a penalty term. Accordingly, we construct the following unconstrained penalized surrogate objective for the original constrained problem.
To effectively solve this reformulated problem, we develop a predictive beamforming framework comprising two distinct architectures: the cross-attention GRU predictive beamforming network (C-GRU-PB-Net) and the cross-attention predictive beamforming network (C-PB-Net), where we first introduce C-GRU-PB-Net as the general architecture and present C-PB-Net as its simplified special case.

\subsection{Proposed C-GRU-PB-Net Framework}
Using the penalty method \cite{liu2022learning,zhang2024transformer}, we reformulate the optimization problem in \eqref{eq:reformulated opt1} as
\begin{align}
\nonumber
    &\mathop{{\text{maximize}}}_{{\bf F}_{n}} \;\;
   {\mathbb{E}}_{{\bf H}_n|{\bf  E}_n^{\tau},{\bf F}_n^{\tau}}\left[\sum_{k = 1}^{K}\log_2\left(\!1+\frac{|{\bf h}_{k,n}^{\sf H}{\bf f}_{k,n}|^2}{\sum_{i\neq k}^{K}|{\bf h}_{k,n}^{\sf H}{\bf f}_{i,n}|^2+\frac{\sigma^2}{P}}\right)\!\right]
   \\
   &\qquad\qquad\;\;-\lambda_1\! \sum_{k = 1}^{K}\!\left(\min\!\left(0, {\frac{\xi_{k,n}({\bf F}_n)}{{\Gamma}_{k,n}}-1}\right)\!\right)^2\!\!\!,
   \label{eq:unconstraintproblem}
\end{align}
where $\lambda_1$ represents the penalty parameter which determines the magnitude of the penalty term.
{The transmit-power penalty is removed by normalizing the network output, which guarantees \eqref{eq:TXpow} while fully utilizing the available power to improve both the communication SE and
sensing SINR.}
To effectively optimize this reformulated problem, we propose a DL architecture designed to directly predict the  beamforming matrix by  processing sensing echoes and historical beamformers as distinct input modalities.
The specific structure of the proposed architecture is shown in Fig.~\ref{fig:systemmodel} and detailed as follows.

\subsubsection{Input Layer}
The complex-valued input is decomposed into its real and imaginary parts and represented as separate features $ \bar{\bf  E}_{n}^{\tau}={\mathcal T}\!\bigl(\mathrm{Re}\{{\bf  E}_{n}^{\tau}\},\mathrm{Im}\{{\bf  E}_{n}^{\tau}\}\bigr)$ and $\bar{\mathbf F}_{n}^{\tau}={\mathcal M}\!\bigl(\mathrm{Re}\{{\bf F}_n^{\tau}\},\mathrm{Im}\{{\bf F}_n^{\tau}\}\bigr)$.
Here, ${\mathcal{T}}(\cdot): {\mathbb{C}^{B \times  K\times L \times  \tau}}\xrightarrow{}{\mathbb{R}^{B\times  2K \times L \times \tau}}$ and ${\mathcal{M}}(\cdot): {\mathbb{C}^{B\times N_t\times K \times \tau  }}\xrightarrow{}{\mathbb{R}^{B\times 2N_t\times K\times \tau }}$ represent the mapping functions, where  $B$ is  the batch size.

\subsubsection{CNN Module}
To extract spatial features from the partitioned inputs, $ \bar{\bf  E}_{n}^{\tau}$ and $\bar{\mathbf F}_{n}^{\tau}$  are fed into separate convolutional neural network (CNN) modules.
Each module consists of 1-D convolutional layers and a ReLU activation function, denoted by $\varphi(x) = {\mathrm{max}(0, x)}$.
The extracted features of the echo signals and beamformers are given by ${\mathbf{\tilde E}}_{n}^{\tau} = \varphi(\text{CNN}({\mathbf{\bar E}}_{n}^{\tau}))$ and $\mathbf{{\tilde F}}_{n}^{\tau} = \varphi(\text{CNN}(\mathbf{{\bar F}}_{n}^{\tau}))$, respectively.

\subsubsection{Cross-attention Module}
To effectively integrate  the separate modalities of echo and beamformer features, we employ  multi-head cross-attention.
{The feature maps extracted from the echo are linearly projected to form the
query matrix ${\bf Q}^{e}$, while the beamformer features are used to form the
key and value matrices ${\bf K}^{b}$ and ${\bf V}^{b}$, respectively, as
\begin{align}
    \label{eq:projection}
    {\bf Q}^{e}
    ={\mathbf{\tilde E}}_{n}^{\tau}{\bf W}_Q,
    {\bf K}^{b}
    ={\mathbf{\tilde F}}_{n}^{\tau}{\bf W}_K,
    {\bf V}^{b}
    ={\mathbf{\tilde F}}_{n}^{\tau}{\bf W}_V,
\end{align}
where
${\mathbf{\tilde E}}_{n}^{\tau}\in\bbR^{N_{\rm e}\times d_{\rm e}}$ and
${\mathbf{\tilde F}}_{n}^{\tau}\in\bbR^{N_{\rm b}\times d_{\rm b}}$,
and the learnable projection matrices satisfy
${\bf W}_Q\in\bbR^{d_{\rm e}\times Hd_{\sf att}}$ and
${\bf W}_K,{\bf W}_V\in\bbR^{d_{\rm b}\times Hd_{\sf att}}$.
Thus, all projections share the same per-head dimension $d_{\sf att}$ for
$H$ attention heads, allowing the query and key features to be multiplied
despite the different token numbers $N_{\rm e}$ and $N_{\rm b}$.}
Subsequently, scaled dot-product attention is used to compute the attention output as
\begin{align}
    \label{eq:attention}
    {\mathrm{Score}({\bf Q}^{e},{\bf K}^{b},{\bf V}^{b})} = \mathrm{softmax}\left(\frac{{\bf Q}^{e}({\bf K}^{b})^{\sf T}}{\sqrt{d_{\sf att}}}\right){\bf V}^{b},
\end{align}
where $\mathrm{softmax}(\cdot)$ denotes the normalization operation  and $d_{\sf att}$ denotes the dimensionality of each attention head.

\begin{algorithm} [t]
\caption{Proposed  Predictive Beamforming} \label{alg:pro}
{\bf{Initialize}}: $t_1 = 0$, random weights $\omega$, {initial penalty weight $\lambda_1^{(0)}$}, and maximum number of iterations $N_{\sf max}$\\
{\bf{ Offline Training:}}
\\
{\hspace*{1em}\bf{Input:}}
Simulated training trajectories and initial beamformers
\\
\hspace*{2em}\While {$t_1$ $<$ $N_{\sf max}$}{\hspace*{2em}Update $\omega$ to minimize $Loss(\omega)$ in \eqref{eq:Lossfunction}\\
\hspace*{2em}{Update $\lambda_1$ using \eqref{eq:lambda_update}}\\
\hspace*{2em}$t_1 \leftarrow t_1+1$}
{\hspace*{1em}\bf{Output:}}{ Well-trained $f_{\omega^{\star}}(\cdot)$}
\\
{\bf{Online Inference:}}
\\
{\hspace*{1em}\bf{Input:}}
{Initial beamformer ${\bf F}_0$ from beam acquisition and} real-time input sequence $(\ddot{\bf  E}_{n}^{\tau}, \ddot{\mathbf F}_{n}^{\tau})$
\\
\hspace*{2em}{\bf{do}} {Beamforming prediction  using $f_{\omega^{\star}}(\cdot)$ }
\\
{\hspace*{1em}\bf{Output: }}{Predicted beamforming matrix ${\bf F}^{\star}_n$ }
\end{algorithm}

\subsubsection{GRU Module}
In the GRU module, a single GRU unit is recurrently applied to the past $\tau$ time slots, i.e., from time slot $(n-\tau)$ to $(n-1)$.
The outputs of the cross-attention module at each time slot are used as the inputs to the GRU in order to capture the temporal correlation of the attention scores, and the hidden state at time slot $(n-1)$ is used as the output of the GRU module, as it captures the temporal dependencies of the historical echo signals and beamformers over the past $\tau$ time slots.

\subsubsection{FC Layer}
To leverage the features extracted by the GRU, its output is fed into a fully connected (FC) layer with linear activation to generate the desired output.
Finally, the optimized beamforming matrix predicted by the proposed algorithm is given by
\begin{align}
    \label{eq:output}
    {{\mathbf F}_{n}=\frac{{\tilde{\mathbf F}}_{n}}{\sqrt{{\rm Tr}({\tilde{\mathbf F}}_{n}{\tilde{\mathbf F}}_{n}^{\sf H})}},}  \quad
    {\tilde{\mathbf F}}_{n}={\mathcal C}({f}_{\omega}(\bar{\bf  E}_{n}^{\tau}, \bar{\mathbf F}_{n}^{\tau})).
\end{align}
Here, $f_{\omega}(\cdot)$ denotes the non-linear transformation of the proposed network parameterized by $\omega$ and  $\mathcal C(\cdot)$ represents the mapping function that constructs the complex-valued matrix.

\begin{remark}[Role of Separate Processing in Cross-Attention and GRU]\normalfont
Echo features act as the query to reflect mobility cues such as Doppler frequency and time delay, which describe vehicle motion, whereas beamformer features serve as the key and value to represent past transmission decisions.
Separate processing of these modalities preserves their modality-specific characteristics before fusion, so that the attention scores indicate how mobility-related echo features align with prior beam decisions.
Furthermore, robustness under high mobility is enhanced by the GRU, which combines these attention-based features over multiple time steps and captures long-term dynamics.
\label{Crossattention}
\end{remark}

\subsection{Proposed  C-GRU-PB-Net Algorithm}
The proposed method operates in two phases: offline training and online inference. For training, the model learns the mapping from sensing echo and beamforming sequences.
For inference, the trained model processes a sequence of sensing echoes and beamforming pairs collected over the previous $\tau$ time slots, and directly predicts the beamforming matrix.

\subsubsection{Offline Training Phase}
The proposed network is trained using mini-batches of $B$ trajectory samples generated by the offline simulator.
For the $j$th sample, we denote the historical sensing-echo and beamformer sequences available at prediction slot $n$ by $(\hat{\bf E}_{n}^{\tau,(j)},\hat{\bf F}_{n}^{\tau,(j)})$.
{Starting from the initial beamformer, the historical beamformers are sequentially generated by the network, and the corresponding sensing echoes are generated according to the sensing signal model.
The training is solver-label-free, i.e., no ground-truth beamformer
obtained from an optimization solver is required. The simulator
generates the vehicle trajectory and the corresponding channel
and sensing parameters, which are used to evaluate the communication
and sensing performance of the predicted beamformer.}
Accordingly, the training loss is expressed as
\begin{align}
    \nonumber
    &Loss(\omega)    =\!-\frac1B\!\sum_{j=1}^{B}\!\sum_{k=1}^{K}\!\log_2\!\left(1+\frac{|({\bf h}_{k,n}^{(j)})^{\sf H}{\bf f}_{k,n}^{(j)}(\omega)|^2}{\sum_{i\neq k}^{K}|({\bf h}^{(j)}_{k,n})^{\sf H}{\bf f}_{i,n}^{(j)}(\omega)|^2+\sigma^2/P}\right)
    \\
   &\qquad\quad+\lambda_1\frac{1}{B}\sum_{j=1}^{B}\sum_{k=1}^{K} \left(\min\!\left(0, {\frac{\xi_{k,n}({\bf F}^{(j)}_n(\omega))}{\Gamma_{k,n}}-1}\right)\right)^2\!\!\!,
   \label{eq:Lossfunction}
\end{align}
where ${\bf f}_{k,n}^{(j)}(\omega)$ denotes the $k$th column of  ${\bf F}_n^{(j)}(\omega)$ for the $j$th training sample.
{Since $\mathbf{F}_n^{(j)}(\omega)$ is the normalized output, the training naturally accounts for the power normalization.}
The network parameters ${\omega}$ are updated iteratively via the backpropagation algorithm to minimize the loss function.
{In addition, $\lambda_1$ is adaptively updated according to the sensing
performance  as
\begin{align}
    \label{eq:lambda_update}
    \lambda_1^{(t+1)}
    =
    \lambda_1^{(t)}
    +\lambda_1^{(0)}\bar g^{(t)},
\end{align}
where
$\bar g^{(t)}
=
\frac{1}{BK}\sum_{j=1}^{B}\sum_{k=1}^{K}
\max\left(0,1-\frac{\xi_{k,n}^{(j)}}{\Gamma_{k,n}}\right)$
denotes the average sensing violation and
$t$ denotes the training iteration.}
Consequently, with the trained network parameters $\omega^{\star}$, the trained model can be represented as $f_{\omega^{\star}}\left(\hat{\bf  E}_{n}^{\tau}, \hat{\mathbf F}_{n}^{\tau}\right)$.

\subsubsection{Online Inference}
{At the initial time slot, the initial beamformer ${\bf F}_0$ is obtained through beam acquisition.
Until $\tau$ echo-beamformer pairs are accumulated, the first $\tau-1$ beamformers are generated with the GRU bypassed, after which the latest $\tau$ pairs are used through a sliding window.
Hence, except for ${\bf F}_0$, the historical beamformers are the previous outputs of the trained network rather than ideal reference beamformers.
The trained model processes these real-time input sequences $(\ddot{\mathbf E}_n^{\tau}, \ddot{\mathbf F}_n^{\tau})$ to generate the predictive beamforming matrix as
\begin{align}
    {\mathbf F}^{\star}_{n}
    =
    \frac{\tilde{\mathbf F}^{\star}_{n}}
    {\sqrt{\mathrm{Tr}\!\left(
    \tilde{\mathbf F}^{\star}_{n}
    \tilde{\mathbf F}_{n}^{\star\sf H}
    \right)}}, \quad
    \tilde{\mathbf F}^{\star}_{n}
    =
    {\mathcal C}\!\left(
    {f}_{\omega^{\star}}
    (\ddot{\bf E}_{n}^{\tau},\ddot{\mathbf F}_{n}^{\tau})
    \right).
\end{align}
where the learned parameters $\omega^{\star}$ are fixed during inference.
Since the BS acquires the sensing echo of every transmission from the physical environment, each prediction reflects the current vehicle state, which helps mitigate the accumulation of prediction errors.
Accordingly, beam acquisition is not required at every time slot and needs to be repeated only when persistent sensing degradation indicates loss of tracking.}
The overall proposed algorithm is summarized in Algorithm~\ref{alg:pro}.

\subsection{Special Case: C-PB-Net}
C-PB-Net retains the separate feature extraction branches and the cross-attention module of C-GRU-PB-Net, while using only the sensing echo and beamformer from the previous time slot $(n-1)$ as  inputs.
It thus removes the GRU-based temporal aggregation and predicts the beamforming matrix for the subsequent time slot based on the fused feature of the previous-slot echo-beamformer pair, thereby providing a lower-complexity alternative to C-GRU-PB-Net.

\begin{figure}[!t]\centering
 \subfigure[Communications performance] {\resizebox{0.7\columnwidth}{!}{\includegraphics{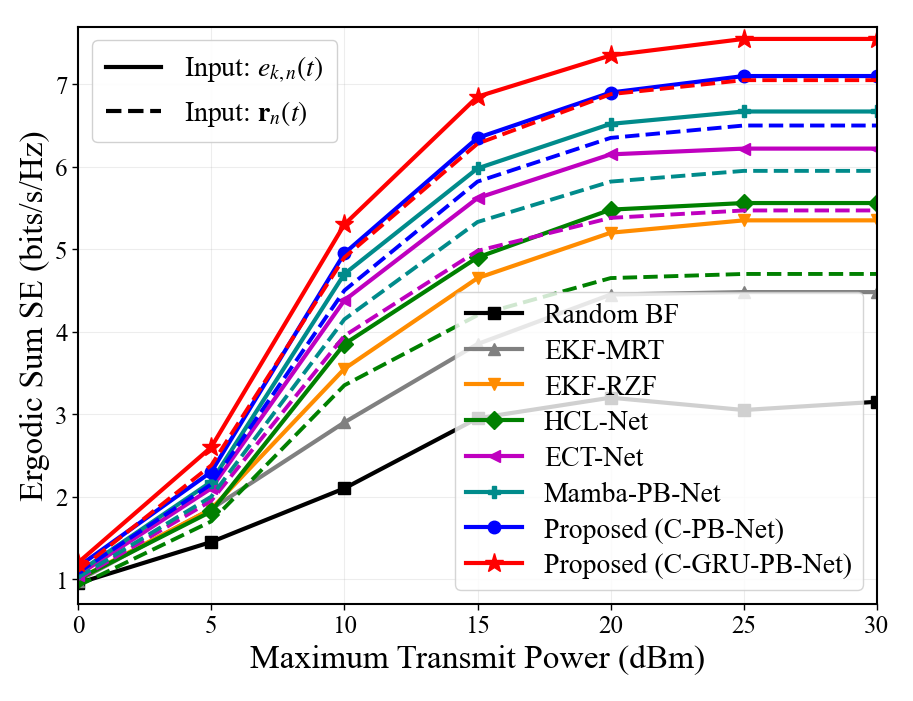}}}
\subfigure[Sensing performance]{\resizebox{0.7\columnwidth}{!}{\includegraphics{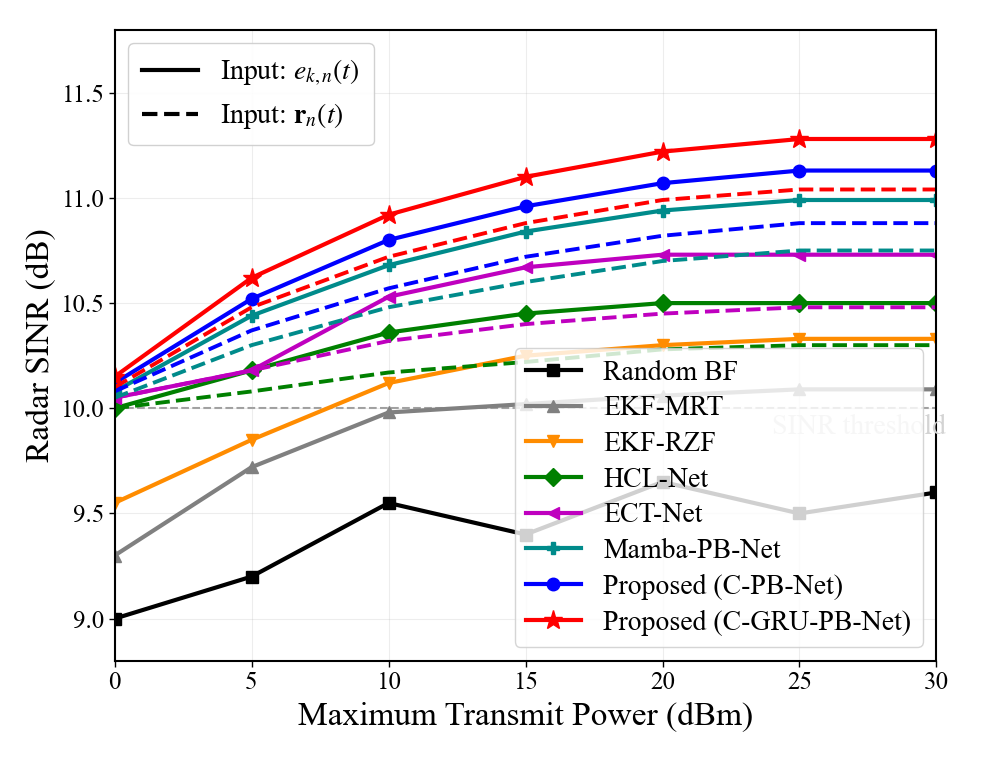}}}
\subfigure[Attention score (Proposed vs. ECT-Net)]{\resizebox{0.9\columnwidth}{!}{\includegraphics{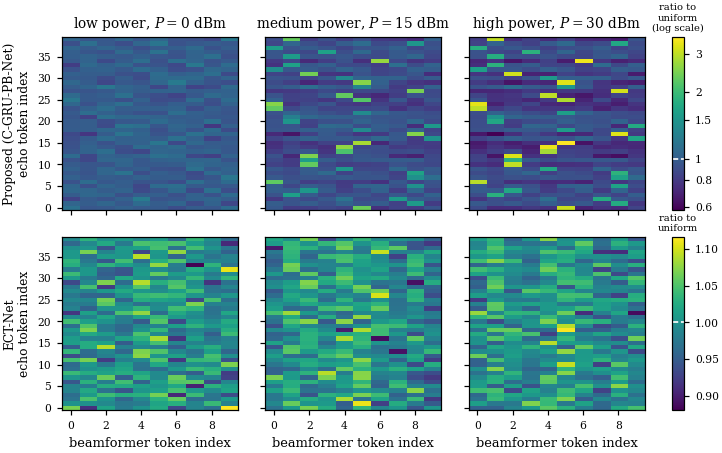}}}
\vspace{-0.5em}
\caption{
 	(a) The  sum SE versus the maximum transmit power, (b) sensing SINR versus the maximum transmit power, and (c) the attention score between echo and beamformer {at low, medium, and high transmit powers}.}
 	\label{fig:sumrate}
\end{figure}

\begin{figure}[!t]\centering
   \subfigure{\resizebox{0.7\columnwidth}{!}{\includegraphics{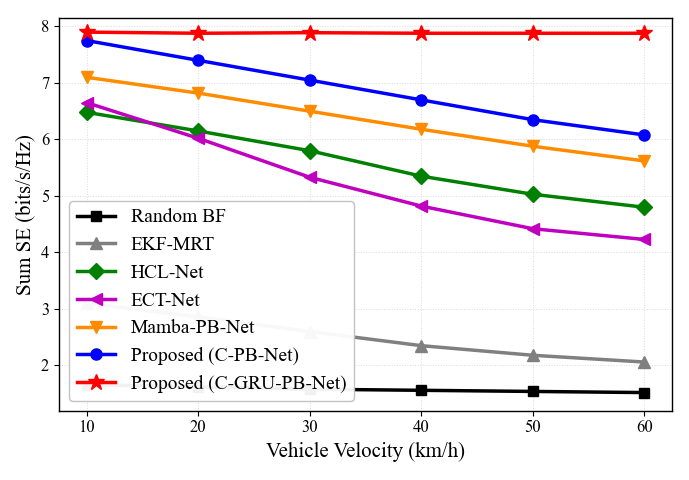}}}
   \vspace{-0.5em}
   \caption{{Sum SE versus the vehicle velocity.}}
   \label{fig:velocity}
\end{figure}

\begin{figure}[!t]\centering
 \subfigure[Communications performance] {\resizebox{0.7\columnwidth}{!}{\includegraphics{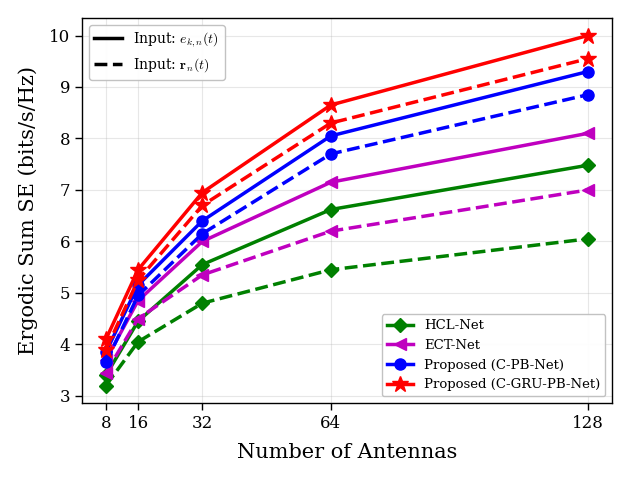}}}
\subfigure[Sensing performance]{\resizebox{0.7\columnwidth}{!}{\includegraphics{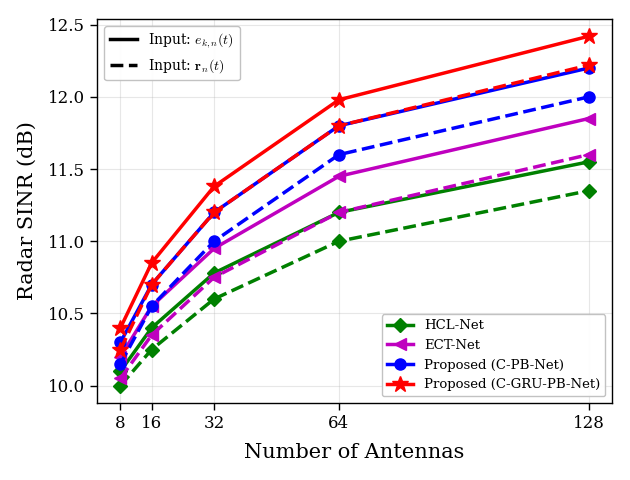}}}
\vspace{-0.5em}
\caption{{Ergodic sum SE and sensing SINR versus the number of antennas at $K = 4$ and $P = 25$ dBm.}}
 	\label{fig:array}
\end{figure}


\section{Simulation results}
The path loss exponent, carrier frequency, complex radar cross-section, and number of past time slots are set to $\eta = 2$, $f_c = 10$ GHz, $\phi = 10 + 10j$, $\lambda_1^{(0)}=500$, and $\tau = 4$, respectively.
We placed the transmit BS at the origin.
The BS-to-user vehicle distances are within $50$ m, and the vehicle velocities range from $50$ km/h to $70$ km/h.
For training, the batch size is set to $B = 64$, and the dataset consists of 4096 training samples and 512 test samples, where the test samples are used to evaluate online inference in the simulation environment.
{We considered the following benchmarks for comparison: 1) Random beamforming (RBF), 2) EKF-maximum ratio transmission (EKF-MRT) \cite{liu2020radar}, 3) EKF-regularized zero-forcing (EKF-RZF), 4) LSTM-based network (HCL-Net) \cite{liu2022learning},  5) Self-attention transformer network (ECT-Net) \cite{zhang2024transformer}, and 6) Mamba-based predictive beamforming network (Mamba-PB-Net), which replaces the cross-attention and GRU modules by a selective state-space model \cite{shi2026channelmamba}}.

In Fig.~\ref{fig:sumrate}, we set $N_t=32$, $N_r=32$, $K=4$, and
$\Gamma_{k,n}=10$ dB.
Fig.~\ref{fig:sumrate}(a) and (b) show the sum SE and sensing SINR,
respectively, while Fig.~\ref{fig:sumrate}(c) shows the attention scores
between sensing echoes and beamformers from previous time slots.
As shown in Fig.~\ref{fig:sumrate}(a) and (b), the proposed
C-GRU-PB-Net/C-PB-Net achieves higher sum SE and sensing SINR than the
considered baselines while satisfying the sensing SINR constraint.
Mamba-PB-Net improves upon HCL-Net and ECT-Net by effectively capturing
temporal dependencies through the selective state-space model, while
C-GRU-PB-Net further improves upon C-PB-Net by exploiting temporal
correlations through the GRU.
In practice, constructing the spatially filtered echo $e_{k,n}(t)$ in
\eqref{eq:rxecho} requires additional preprocessing such as angle estimation
and spatial filtering.
We therefore also consider the raw received echo ${\bf r}_n(t)$ as the network
input.
Although using ${\bf r}_n(t)$ degrades the overall communication and sensing
performance, the proposed C-GRU-PB-Net exhibits a smaller degradation than the
learning-based baselines and maintains its performance advantage.
As shown in Fig.~\ref{fig:sumrate}(c), compared with ECT-Net, the proposed
network exhibits more structured attention patterns that better capture the
echo-beamformer correlation.
Moreover, its attention map becomes more concentrated as the transmit power
increases, reflecting the improved reliability of the sensing echo, whereas
ECT-Net shows less variation.
Overall, these results demonstrate the effectiveness and robustness of the
proposed C-GRU-PB-Net in exploiting echo-beamformer relationships for predictive beamforming.


{In Fig.~\ref{fig:velocity}, we evaluate the effect of the vehicle mobility by sweeping the velocity from $10$ to $60$ km/h.
As the velocity increases, the channel varies more rapidly between consecutive slots, which makes beam prediction more challenging.
The learning-based baselines and C-PB-Net gradually degrade, whereas C-GRU-PB-Net maintains relatively stable performance owing to the GRU module that captures the temporal dependencies across multiple time slots, so that its advantage becomes more pronounced at higher velocities.}

{In Fig.~\ref{fig:array}, we plot the communication and sensing performance with
respect to the number of antennas by considering
$N_t=N_r\in\{8,16,32,64,128\}$.
Both the communication and sensing performance improve as the number of antennas
increases, since a larger transmit array provides increased array gain, while a
larger receive array improves spatial resolution and reduces the residual
inter-beam leakage in \eqref{eq:rxecho}.
Across all considered array sizes, the proposed C-GRU-PB-Net consistently
achieves the highest sum SE while satisfying the sensing threshold, and exhibits
a smaller degradation than the learning-based baselines when the raw echo is
directly used as input.
These results indicate that the proposed framework is well suited for future
6G systems employing massive MIMO \cite{isac2026fr3}.}

Finally, we evaluate the computational complexity in terms of floating-point operations (FLOPs), which quantify the number of arithmetic operations required for inference.
Using the default hyperparameters in our implementation, the estimated inference complexity per sample is approximately 2.71 MFLOPs for ECT-Net, 2.62 MFLOPs for C-PB-Net, and 3.12 MFLOPs for C-GRU-PB-Net.
Hence, C-PB-Net achieves both lower inference complexity and higher performance than ECT-Net, and C-GRU-PB-Net further improves the performance with only a modest increase in FLOPs.

\section{Conclusion}
In this paper, we proposed a DL-based predictive beamforming framework for vehicular ISAC networks that explicitly decouples sensing echoes and historical beamformers to preserve their modality-specific characteristics.
We formulated a sum SE maximization problem under a sensing SINR constraint and converted it into a DL training loss function.
The decoupled features are then fused through cross-attention, while GRUs capture their temporal evolution across multiple time slots.
Simulation results confirmed that the proposed framework outperforms the considered model-based and learning-based baselines while satisfying the sensing requirement.
The proposed framework therefore provides a practical predictive beamforming solution with reduced pilot overhead and low inference complexity for connected and autonomous vehicular networks.

\bibliographystyle{IEEEtran}
\bibliography{Ref}
\vspace{12pt}

\end{document}